# Measurement of the Optical Conductivity of Graphene

Kin Fai Mak[1], Matthew Y. Sfeir[2], Yang Wu[1], Chun Hung Lui[1], James A. Misewich[2] and Tony F. Heinz[1*]

[1]Departments of Physics and Electrical Engineering, Columbia University, 538 West 120$^{th}$ St., New York, NY 10027, USA
[2] Brookhaven National Laboratory, Upton, NY 11973, USA

## Abstract

Optical reflectivity and transmission measurements over photon energies between 0.2 and 1.2 eV were performed on single-crystal graphene samples on a $SiO_2$ substrate. For photon energies above 0.5 eV, graphene yielded a spectrally flat optical absorbance of (2.3 ± 0.2)%. This result is in agreement with a constant absorbance of $\pi\alpha$, or a sheet conductivity of $\pi e^2/2h$, predicted within a model of non-interacting massless Dirac Fermions. This simple result breaks down at lower photon energies, where both spectral and sample-to-sample variations were observed. This "non-universal" behavior is explained by including the effects of doping and finite temperature, as well as contributions from intraband transitions.

PACS numbers: 71.20.Tx, 78.20.Ci, 78.30.Na, 78.66.Tr

Graphene, a single-monolayer of $sp^2$-hybridized carbon, has been predicted to exhibit a particularly simple optical absorption spectrum. In the infrared-to-visible spectral range, the expected absorbance has been calculated to be independent of frequency and to have a magnitude given by $\pi\alpha$ = 2.293% [1-8], where $\alpha = e^2/\hbar c$ denotes the fine-structure constant (in cgs units). The prediction of such a universal absorbance, one not affected by the details of the graphene band structure, is equivalent to an optical sheet conductivity for graphene exhibiting a frequency-independent value of $\sigma = \pi G_0/4$, where $G_0 = 2e^2/h$ is the quantum of conductance [9]. Many of the electronic properties of graphene, with its linear dispersion near the Fermi energy, can be modeled in terms of massless Dirac Fermions. In this context, the predicted conductivity can be viewed as an intrinsic property of two-dimensional massless Fermions. It arises from a chiral resonance [10], a process in which a particle-antiparticle pair is created upon the absorption of a photon, and the result for graphene follows directly upon consideration of its four-fold spin-nodal degeneracy [10, 11]. This universal optical conductivity may be viewed as the high-frequency counterpart of the minimal dc conductivity of graphene, a subject of much recent attention. The minimal dc conductivity has also been measured experimentally to have a value on the order of $G_0$ [12, 13]. However, the universality of the value is still under debate due to its sensitivity to the local sample environment. In contrast, although the relative change of the optical sheet conductivity of graphene at different charge densities has been demonstrated recently in the mid-infrared range [14, 15], for higher photon energies we expect its value to exhibit little sensitivity to the local environment of the sample and to exhibit universal behavior [1-4].

In this letter, we present an experimental determination of the optical conductivity of graphene over the spectral range of 0.2 – 1.2 eV. The results were obtained by direct measurement of the optical transmission and reflection of large-area, single-crystal graphene samples prepared on a transparent substrate. Over the spectral range of 0.5 – 1.2 eV, we observe a frequency-independent absorbance of $A$ = (2.3 ± 0.2)% = (1.0 ± 0.1) $\pi\alpha$ or sheet conductivity of $\sigma$ = (6.1 ± 0.6) x $10^{-5}$ S = (1.0 ± 0.1) $\pi G_0/4$, in accordance with the idealized model of interband transitions [16]. The results were reproduced for different graphene samples, showing the optical conductivity over the corresponding spectral range to be a robust quantity. A clear breakdown of the universal behavior was, however, observed when measurements were extended down to photon energies of 0.2 eV. For these lower photon energies, significant spectral variation was observed in the absorption, as well as differences in the response exhibited by different samples. We can account for these departures from the idealized result considering of the effects of finite temperature and a doping-induced shift of the chemical potential from the charge-neutrality (Dirac) point. Thus in addition to verifying the ideal, universal behavior of the optical conductivity in graphene over the relevant spectral range, in this paper, we identify limitations imposed by finite temperature and doping.

In our measurements we investigated graphene samples supported on bulk $SiO_2$ substrates (Chemglass, Inc). After carefully cleaning the substrates by sonication in methanol, the graphene samples were deposited by mechanical exfoliation [17] of Kish graphite. In addition to the use of the optical absorption spectra described below, the Raman peak associated with the 2D feature was employed to identify regions of the substrate containing single-layer graphene crystals [18]. The area of these regions varied from several hundreds to thousands of μm².

The optical absorption of graphene was determined in both the transmission and reflection geometry in the near-infrared spectral region using the U2B beamline of the National Synchrotron Light Source at Brookhaven National Laboratory as a source of bright broadband radiation. The optical radiation passing through or reflected by the sample was detected with a Nicolet micro-FTIR spectrometer equipped with an MgCdTe detector under nitrogen purge. Using a 32× reflective objective, the synchrotron radiation could be focused to a spot size of 10 μm. The absorption and reflectance spectra of the graphene were obtained by normalizing the sample spectrum with that from the bare substrate. As a cross-check on the absolute magnitude of the signal, we performed independent measurements of the optical response near the upper end of the 0.5 – 1.2 eV spectra using a conventional tungsten halogen lamp as the optical source.

Fig. 1 shows the measured reflection spectrum for graphene on the substrate ($R_{g+s}$) compared with that of bare substrate ($R_s$). The fractional change in the reflectance (right scale of Fig. 1) is about 9% and exhibits no significant frequency dependence. This quantity is directly related to the absorbance $A$ of the graphene film. For a layer of material like graphene of thickness $d << \lambda$, the wavelength of light, that is supported by a thick transparent substrate, the fractional change of reflectance obtained by solving the Maxwell's equations is [19]

$$\delta_R = \frac{R_{g+s} - R_s}{R_s} = \frac{4}{n_s^2 - 1} A \quad , \tag{1}$$

where $n_s$ is the refractive index of the underlying SiO$_2$ substrate. We have assumed here normal incidence, corresponding to our experimental conditions. Since the light is focused on the sample (numerical aperture NA< 0.5), we have also considered the possible influence of the spatial Fourier components of the beam away from normal incidence. By generalizing Eqn. (1) to finite angles of incidence, we find that the influence of the beam focusing on the change in reflectance for unpolarized light over the relevant range of incidence angles is very slight (~ 1% correction). This result was confirmed experimentally by demonstrating that no meaningful difference was observed in the absorbance measured with objectives having different NAs.

An analogous expression to Eqn. (1) applies for the case of the transmission geometry, but with the factor of $\frac{4}{n_s^2 - 1}$ replaced by $\frac{2}{n_s + 1}$. In our analysis, we emphasize the reflectance data because the change in the optical response induced by the presence of the graphene layer is several times larger than for transmission, yielding higher measurement accuracy. The transmission results are, however, fully consistent with the reflectance measurements.

Figure 2 shows the frequency-dependent absorbance *A* for three different graphene samples over the spectral range of 0.5 – 1.2 eV. The data are also presented in terms of the (real part of the) sheet conductivity σ of the sample, which for a thin film is related to the absorbance by σ = (c/4π) *A*. In the analysis, we included the slight dispersion (~ 1%) of the substrate refractive index [20].

We would now like to make some observations about the results of the absorption spectra in Fig. 2. (1) Over this spectral range, different samples of monolayer graphene, three of which are shown in the Figure, yield equivalent response. Unlike the behavior for the minimal conductivity, the optical conductivity over this frequency range is not significantly influenced by the detailed nature of the sample or its local environment. (2) The absorbance *A* is spectrally flat within a band of ±10%. This range is somewhat larger than the fluctuations in our measurements, and the data suggest a slight increase of absorbance with photon energy. Possible explanations for this weak departure from a flat absorbance are discussed at the end of the paper. (3) The magnitude of the absorbance *A* is consistent with the universal value of πα = 2.293%, shown as the black horizontal line in the Figure. Computing the average and variance of three spectra in the Figure over the indicated range, we obtain <*A*> = (2.28 ± 0.14)%

In analyzing our experimental results, it is useful to consider the expected behavior if we relax some of the idealizations used to obtain the simple, universal behavior of constant absorbance of πα. Two key factors are the effect of finite temperature and of finite doping, since the measurements were performed at room temperature and our exfoliated graphene samples typically exhibit appreciable spontaneous doping. The principal effect of a finite temperature (*T*) and non-zero chemical potential ($\mu$, relative to the Dirac point) from doping is to reduce the transition strength because of state blocking. This can be taken into account simply by including the Fermi-Dirac distribution in the calculation of the optical conductivity at frequency $\omega$ [4-8]:

$$\sigma(\omega, T) = \frac{\pi e^2}{4h} \left[ \tanh\left(\frac{\hbar\omega + 2\mu}{4k_B T}\right) + \tanh\left(\frac{\hbar\omega - 2\mu}{4k_B T}\right) \right] \quad (2)$$

We see that for photon energies appreciably greater than twice the shift in Fermi energy and thermal energy, there are no state-blocking effects, and we revert to the universal behavior. For realistic experimental parameters, this situation is expected to prevail for photon energies ≥ 0.5 eV, as seen experimentally.

While the interband transitions dominate the response at high photon energies, at lower photon energies intraband transitions also become important. The exact form of the intraband contribution depends on the scattering mechanisms involved and is closely related to the problem of the minimal dc conductivity [3, 4, 6-8]. For our present purpose, a phenomenological scattering rate Γ is assumed and the intraband contribution to the sheet conductivity (with the Fermi level at the Dirac point) can be shown to possess a Drude form [21]:

$$\sigma_{\text{intraband}}(\omega,T) = 4\ln 2 \frac{e^2}{h} \frac{(\hbar\Gamma) k_B T}{(\hbar\omega)^2 + (\hbar\Gamma)^2} \tag{3}$$

This contribution decreases in magnitude with increasing photon energy, but may still be appreciable in the mid-infrared spectral region for typical scattering rates (tens of meV).

The predicted optical conductivity from both inter- and intraband contributions for graphene at room temperature is shown in Fig. 3. The different curves, which correspond to representative values of the chemical potential and scattering rate, were calculated using the Kubo formula to include both Eqn. (2) and the generalization of Eqn. (3) for finite doping [3, 4, 6, 8]. At lower photon energies, we see a significant departure from the universal value for the conductivity. We also observe a sensitive dependence on chemical potential and thus anticipate variation among graphene samples, each of which may exhibit different levels of spontaneous doping.

Figure 4 displays the measured absorption of graphene for lower photon energies. It is immediately apparent that the universal behavior, which prevails above 0.5 eV, does not apply. There is a significant deviation from the value of the universal absorbance. In addition, unlike for higher photon energies, the results differ from sample to sample. The absorbance of sample 1 decreases significantly below 0.4 eV, while sample 2 shows only a modest decrease. This difference is attributed to a higher doping level in sample 1 compared to sample 2. The experimental results are compatible with the prediction of the analysis described above in which we include both the inter- and intraband contributions at finite temperature and doping levels (red lines in Fig. 4).

Finally, we wish to mention the limitations of an analysis, such as that above, based on a model of graphene excitations as those of non-interacting massless Dirac Fermions. First, there are possible *many-body effects*. Coulomb interactions in carbon nanotubes are, for example, known to lead to tightly-bound excitons [22-24]. Such many-body effects have been predicted to lead to a reduction of the optical conductivity at lower photon energies [10]. Second, one must consider the influence of the *deviation from a linear dispersion relation* in the graphene band structure as one moves away from the Dirac point [25]. For photon energies above 2.5 eV, these band-structure effects are expected to increase the optical absorption [5, 25]. In view of the very short lifetime of excited states in graphene [26, 27], significant spectral broadening may be present, leading to enhanced absorption even at lower photon energies. Either or both of these effects may account for the slight upward trend of the graphene absorbance with increasing photon energy that is present in the experimental data of Fig. 2. An unambiguous identification of these interesting effects will, however, require further measurements.

The authors at Columbia University acknowledge support from the Nanoscale Science and Engineering Initiative of the NSF under grant CHE-0117752, the New York State Office of Science, Technology, and Academic Research (NYSTAR), and the US Department of Energy (DOE) under grant DE-FG02-03ER15463; the authors at Brookhaven were supported under US DOE contract DE-AC02-98CH10886. The synchrotron studies were supported by the National Synchrotron Light Source at Brookhaven and the Center for Synchrotron Biosciences, Case Western Reserve University under grant P41-EB-01979 with the National Institute for Biomedical Imaging and Bioengineering. The authors would like to thank Dr. Mikito Koshino for useful discussions.

**Figures:**

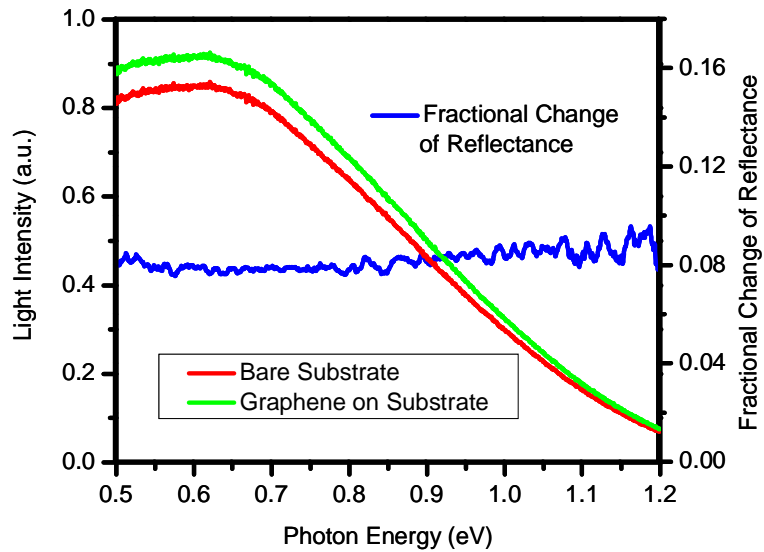

Fig. 1. Experimental data for the reflection from a graphene monolayer on the $SiO_2$ substrate for photon energies in the range of 0.5 – 1.2 eV. The curves associated with the left vertical axis are the measured reflection spectra of the substrate alone (red) and of the graphene monolayer on the substrate (green). The curve associated with the right vertical axis represents the fractional change in reflectance from the graphene monolayer.

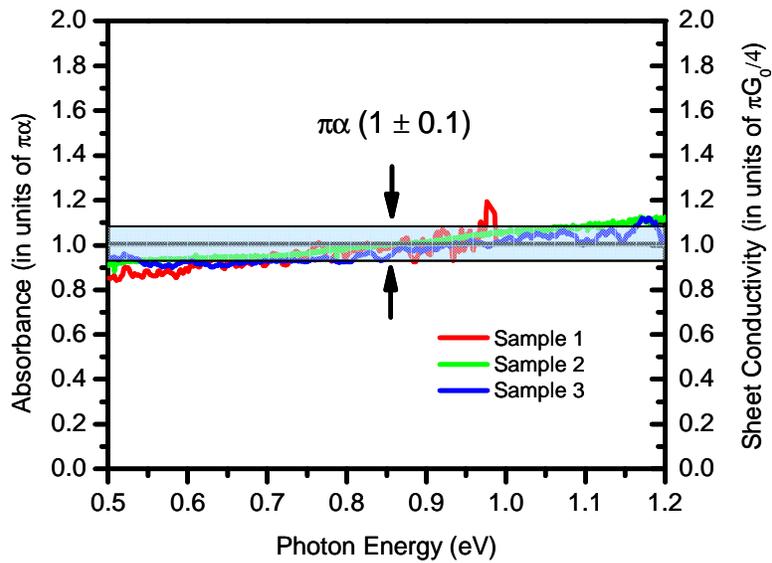

Fig. 2. Absorption spectra for three different samples of graphene over the range of photon energies between 0.5 and 1.2 eV. The left scale gives the absorbance in units of πα, while the right scale gives the corresponding optical sheet conductivity in units of $\frac{\pi}{4}G_0 = 6.08\times10^{-5}$ S. The black horizontal line corresponds to the universal result of an absorbance of πα = 2.293%, with a range indicated of ± 0.1 πα or approximately ± 0.2%.

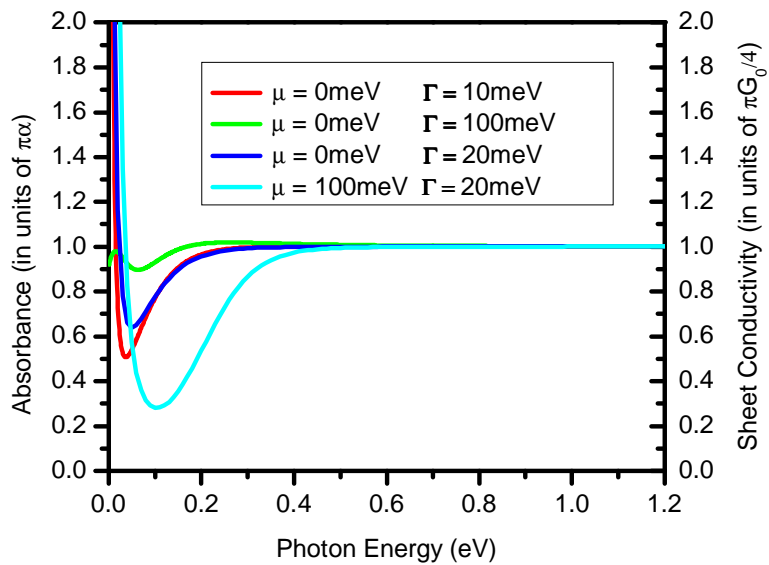

Fig. 3. Calculated graphene absorption spectra from 0 to 1.2 eV, including both the inter- and intraband contributions at a temperature of 300 K. The calculation

is based on a model of non-interacting massless Fermions. Results are shown for different values of the chemical potential $\mu$ and scattering rate $\Gamma$.

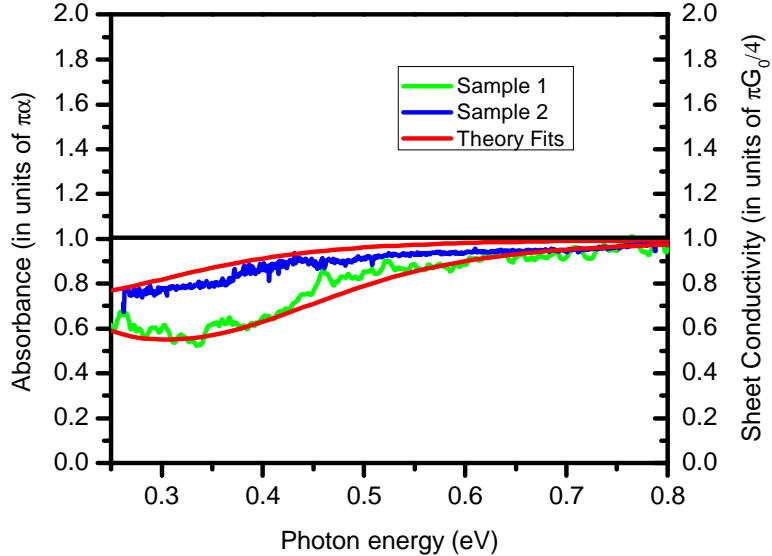

Fig. 4. Measured graphene absorption spectra of samples 1 and 2 over a range of photon energies between 0.25 and 0.8 eV. The smooth red lines are based on the theory shown in Fig. 3, with µ = 200 and 100 meV for samples 1 and 2, respectively.